\documentclass[12pt]{article}

\usepackage{amscd}

\usepackage{verbatim}

\usepackage{amssymb}

\usepackage{amsmath}

\newcommand{\be}{\begin{equation}}
\newcommand{\ee}{\end{equation}}
\newcommand{\ben}{\begin{eqnarray}}
\newcommand{\een}{\end{eqnarray}}
\newcommand{\ba}{\begin{eqnarray}}
\newcommand{\ea}{\end{eqnarray}}

\newcommand{\bi}{\begin{itemize}}
\newcommand{\ei}{\end{itemize}}

\begin{document}

\begin{center}

\vspace{24pt} { \large \bf Stability analysis of the
Witten black hole (cigar soliton) under world-sheet RG flow} \\

\vspace{30pt}

\vspace{30pt}

{\bf Carolyn Lambert} \footnote{carolyn9y@gmail.com}, {\bf Vardarajan
Suneeta$^{(a)}$}\footnote{suneeta@iiserpune.ac.in}

\vspace{24pt} 
{\em  (a) The Indian Institute of Science Education and Research (IISER),\\
Pune, India - 411021.}

\end{center}

\date{\today}
\bigskip

\begin{center}
{\bf Abstract}\\
\end{center}
\vskip 0.8cm
We analyze the stability of the Euclidean Witten black hole (the cigar soliton in mathematics literature) under first-order RG (Ricci) flow of the world-sheet sigma model. This analysis is from the target space point of view. We find that the Witten black hole  has no unstable normalizable perturbative modes in a linearized mode analysis in which we consider circularly symmetric perturbations. Finally, we discuss a result from mathematics that implies the existence of a nonnormalizable mode of the Witten black hole under which the geometry flows to the sausage solution studied by Fateev, Onofri and Zamolodchikov.

\newpage

\section{Introduction}
\setcounter{equation}{0}
Black hole solutions in dimensions less than four are important as tractable models for understanding black hole physics. One such example, the Witten black hole, arises while studying the world-sheet sigma model in closed string theory on a two-dimensional target space. This geometry also arises in mathematics as the fixed point of a flow equation called the Ricci-de Turck flow --- it is termed the `cigar soliton'. In the physics of the world-sheet sigma model, `classical' solutions (exact string backgrounds) are obtained by setting the $\beta$ functions of this theory to zero --- thus, they are fixed points of the renormalization group (RG) flow of the sigma model. The beta functions are obtained in an expansion in powers of an appropriate coupling constant $\alpha'$ (square of string length) in the sigma model. In the case where the fields other than the metric are set to zero, to first order in $\alpha'$, this coincides with the vacuum Einstein equation. Upon considering both the metric and the dilaton and setting the $\beta$ functions to zero, an interesting solution that has been obtained is the metric exterior to the horizon of a two dimensional black hole (see \cite{mandal}, \cite{witten} and references therein). This solution is a fixed point of the first order RG flow of the metric and dilaton (which is incidentally the same as the Ricci de Turck flow of mathematics mentioned above). Due to RG scale invariance, there is a natural expectation that the sigma model with this target space should be a conformal field theory (CFT).  Witten has shown that starting from the $SL(2, R)/U(1)$ gauged Wess-Zumino-Witten (WZW) model, which is an exactly solvable CFT, one could obtain the sigma model corresponding to this Wick-rotated (Euclidean) black hole \cite{witten}. This strongly suggests that the physics of string propagation in this black hole can be inferred from this WZW CFT --- this has been discussed (particularly extending to results valid for the Lorentzian black hole) first in \cite{dvv}.\\

In this paper, we perform a target-space study of the (Euclidean) Witten black hole - specifically, of the stability of this black hole under \emph{world-sheet RG flow}. There are many related motivations for this problem. If the Euclidean Witten black hole is unstable under RG flow --- then the natural question to investigate is the end-point of the flow. Does the flow lead us to another geometry described by a world-sheet CFT, or does it lead to a singular geometry that is indicative of a nontrivial (nonperturbative) IR fixed point? If there are no unstable normalizable modes, then since the black hole geometry is described by a CFT, this analysis would provide a complementary target space picture to the usual analysis of relevant deformations of the CFT.
In a separate section, we also discuss a result from mathematics that could be interpreted as the existence of a particular type of \emph{nonnormalizable} mode of the Witten black hole. We suggest that this mode leads to a geometry change from the black hole to the (uncompactified) sausage solution --- the sausage is an exact solution to RG flow discussed first by Fateev, Onofri and Zamolodchikov \cite{foz}. The sausage solution (called Rosenau solution \cite{rosenau} in math literature) is believed to be described by a sigma model which is integrable. The curvature of this solution increases along RG flow, and thus it has a nontrivial IR limit which needs to be described nonperturbatively. In \cite{foz}, certain scattering theories which are candidates for a nonperturbative IR limit are discussed. It would be interesting to understand the world-sheet interpretation of this unstable mode.\\

There are ideas motivated by closed string field theory, that the RG flow parameter could play the role of a dynamical time (in a certain approximation) in string field theory --- a review of these ideas can be found in \cite{HMT}. Other authors have discussed instabilities of black holes under RG flows and argued using these ideas, that this could be a dynamical instability mediated by tachyons --- see, for instance \cite{barbon} and \cite{hw}. Thus, in the context of these ideas, the issue of stability of the Witten black hole is of interest. One could also consider a gravity-dilaton path integral, where the action is the string low energy effective action in the presence of a dilaton. The Witten black hole is a stationary point of this action. The first variation of this action gives a term that depends on beta functions and vanishes on a stationary point. The second variation of the action, evaluated at the stationary point is crucial, because it determines the stability of the stationary point in the path integral formulation of quantum gravity. This is similar to the analysis of Gross, Perry and Yaffe (GPY)  \cite{GPY} for the Euclidean Schwarzschild instanton. If we find an instability in our analysis, this will probably be indicative of an instability in the path integral due to the fact that we effectively compute the second variation of a part of the action --- one major difference however is that the dilaton is not varied freely. Rather, the dilaton is used to fix conformal gauge. In another context, constraining the dilaton corresponding to a different gauge fixing choice for this action was done in \cite{tseytlin}. Our main result is that we rule out unstable modes when the perturbations are circularly symmetric, which seems to indicate dynamical stability of the black hole in the sense we have mentioned (under such perturbations).\\

Lastly, we would like to mention a related result in mathematics \cite{ds} which proves that the only solution to RG flow that is nonsingular both in the UV and IR and satisfies some other conditions is the Witten black hole. In physics, however, we are not interested only in solutions that are nonsingular both in the UV and IR. In fact, many well-studied quantum field theories either require a UV completion or have nontrivial IR limits under RG flow. Thus this result is not directly of relevance here. A different question, the stability of the Witten black hole-tachyon system due to \emph{on-shell} perturbations (i.e., those solving equations obtained by setting the $\beta$ functions of this system to zero) has been addressed in \cite{marcus}, \cite{peet} and \cite{gs}.\\

In the next section, we briefly introduce the Witten black hole and the RG flow of which it is a fixed point (Ricci-de Turck flow). Section 3 has the main computation of linearized analysis of stability of the Witten black hole and details of gauge fixing. In this section, we show that the Witten black hole is stable under RG flow in a linearized mode analysis for (circularly symmetric) normalizable perturbations. In Section 4, we discuss a result from mathematics that implies the existence of a nonnormalizable mode under which we speculate that the Witten black hole flows to the sausage metric.

\section{An introduction to first-order RG (Ricci) flow and some special solutions}
\setcounter{equation}{0}
The Ricci flow is a partial differential equation in mathematics that describes a flow through Riemannian metrics on a manifold. It was used recently in the proof of the Poincare conjecture \cite{perelman}.  The flow is
\begin{eqnarray}
\frac{\partial \tilde g_{ab}}{\partial \tau} = - \alpha'
\tilde R_{ab}
\label{2.1}
\end{eqnarray}

This flow arises in physics as the simplest lowest-order (in square of string
length $\alpha'$) RG flow of the world-sheet sigma model for closed
strings. In this context, the Ricci flow is the flow of the metric
of the target space with respect to the RG flow parameter $
\tau$.
Both in physics and mathematics, what is of interest is a flow of metrics mod
diffeomorphisms. In the mathematics literature, $\alpha' =
2$. Henceforth, to make contact with existing results in mathematics, we will consider Ricci flow with respect to a parameter $t = \frac{\alpha'}{2} \tau $ which is the `Ricci flow time' in mathematics (this is not the time coordinate of a physical spacetime). The Ricci/RG flow with respect to this parameter is
\begin{eqnarray}
\frac{\partial \tilde g_{ab}}{\partial t} = - 2
\tilde R_{ab}.
\label{2.1a}
\end{eqnarray}

 A metric which changes only by a $t$-dependent diffeomorphism along the flow does not lead to new physics. Therefore all flows related to each other by $t$
dependent diffeomorphisms generated by a vector field $V$ are
equivalent; and it is more useful to consider the generic flow in this class
--- the {\em Ricci-de Turck} flow:
\begin{eqnarray}
\frac{\partial g_{ab}}{\partial t} = - 2
(R_{ab} + \nabla_{a} V_b + \nabla_{b} V_a )
\label{2.2}
\end{eqnarray}

The solution to (\ref{2.2}), $g_{ab}$ is related to the solution to (\ref{2.1}), $\tilde g_{ab}$ by a $t$ dependent diffeomorphism generated by $V$.
This flow describes a flow of geometries on the manifold (target space). When $V_a = \nabla_a \Phi$ (gradient of a function), then the Ricci-de Turck flow is precisely the RG flow of the world-sheet sigma model for a closed string in the presence of a dilaton, the dilaton being  $\Phi$. Fixed points of the flow in this case,  which satisfy
\begin{eqnarray}
R_{ab} + 2 \nabla_{a} \nabla_b \Phi = 0
\label{2.3}
\end{eqnarray}
are `exact string backgrounds', in which the associated sigma model is expected to be described by a conformal field theory (CFT). In the mathematics of Ricci flow, the solutions to (\ref{2.3}) are called {\em gradient steady solitons}. It is known that the only such solitons on compact manifolds are Einstein metrics. On noncompact manifolds, there are many solitons known on $R^{n}$ and $C^{n}$ (see \cite{cao} for an extensive review). The most exciting of these (for physics) is the {\em cigar soliton} or the Euclidean Witten black hole. The Witten black hole is described by a known CFT (the WZW model) and an interesting question is whether the other solitons known can also be described by CFTs.
The Euclidean Witten black hole (cigar soliton) is the following metric on $R^2$ ($r$ and $\theta$ are the usual polar coordinates):
\begin{eqnarray}
ds^2 = \frac{1}{1+r^2} (dr^2 + r^2 d\theta^2).
\label{2.4} \end{eqnarray}
On analytically continuing the metric by letting $\theta = i t$ where $t$ is to be thought of as a time coordinate, this becomes the (Lorentzian) metric of a black hole (as discussed in \cite{witten}). In this paper, we will be concerned only with the metric (\ref{2.4}). This metric is asymptotically cylindrical as $r \rightarrow \infty$. It solves (\ref{2.3}) with the dilaton $\Phi = - \frac{1}{2} \ln (1 + r^2 )$ (we can add an arbitrary constant to the dilaton --- the exponential of this constant is related to the mass of the corresponding Lorentzian black hole). The metric (\ref{2.4}) is a fixed point of the flow (\ref{2.2}) with $V_a = \nabla_a \Phi$ (and $\Phi$ as above), but is also a solution to the Ricci flow (\ref{2.1a}) or (\ref{2.1}) after the diffeomorphism $ \tilde r = e^{-2t} r$. This solution is
\begin{eqnarray}
ds^2 = \frac{1}{e^{4 t} +r^2} (dr^2 + r^2 d\theta^2).
\label{2.5} \end{eqnarray}
So from the point of view of geometry, the gradient of the dilaton merely generates $t$ dependent diffeomorphisms.
\section{Linearized stability analysis of the cigar soliton/Euclidean Witten black hole}

Let us consider a small perturbation of the Witten black hole metric, the perturbed metric being $g_{ab}^{p} = g_{ab}^{Witten} + \epsilon h_{ab}$,
where $0 < \epsilon < 1$ is a parameter indicating the smallness of the perturbation. We would like to study the evolution of the perturbation under RG flow. We set up the linearized stability problem below and demonstrate stability of the black hole under a wide class of perturbations. We can rewrite all the results of this section in terms of the RG flow parameter $\tau$ by using $t = \frac{\alpha'}{2} \tau $.

Since the metric is two-dimensional, without loss of generality, we can choose a gauge so that the perturbed metric $g_{ab}^{p}$ is conformal to the flat metric. We consider circularly symmetric perturbations, and the corresponding perturbed metric is the initial condition for the flow (\ref{2.2}). With a suitable choice of gauge ($V$), we can easily see that the flow preserves isometries. By choosing gauge appropriately (conformal gauge), the evolving perturbed metric is of the form
\begin{eqnarray}
ds^2 = G(r, t) (dr^2 + r^2 d\theta^2).
\label{3.3} \end{eqnarray}
where $G(r) = [\frac{1}{1+r^2} + \epsilon F(r, t)]$.

One motivation for choosing the perturbations to be $\theta$ independent is that on Lorentzian continuation, $\theta$ becomes the Lorentzian time and these would correspond to static perturbations. But from the computational point of view, the gauge fixing procedure we do below becomes much harder when perturbations are $\theta$ dependent.

The right gauge choice is found as follows: Consider the flow (\ref{2.2}) with $V_a = \nabla_a \Phi$,
\begin{eqnarray}
\frac{\partial g_{ab}}{\partial t} = - 2
(R_{ab} + 2 \nabla_{a} \nabla_{b} \Phi )
\label{3.4}
\end{eqnarray}

For the metric (\ref{3.3}), $g_{r \theta} = 0 $ and $g_{\theta \theta} = r^2 g_{r r}$. Recall that in two dimensions, $R_{ab} = \frac{1}{2} g_{ab} R$
where $R$ is the scalar curvature. For consistency of (\ref{3.3}) as a solution to (\ref{3.4}), we must therefore first satisfy  $\nabla_{r} \nabla_{\theta} \Phi = \nabla_{\theta} \nabla_{r} \Phi = 0$. If we consider $\Phi(r, t)$ to be independent of $\theta$, then for the metric (\ref{3.3}), $\nabla_{r} \nabla_{\theta} \Phi = \nabla_{r} \partial_{\theta} \Phi = 0$.
\begin{eqnarray}
\Gamma_{r \theta}^{r} = \frac{1}{2 G(r, t)} \frac{\partial G(r, t)}{\partial \theta} = 0.
\label{3.5}
\end{eqnarray}

Then it is easy to see that $\nabla_{\theta} \nabla_{r} \Phi = 0$. For consistency of (\ref{3.3}) as a solution to (\ref{3.4}), we also need
$\nabla_{\theta} \partial_{\theta} \Phi = r^2 \nabla_{r} \partial_{r} \Phi $. For the metric (\ref{3.3}), this leads to the following equation:
\begin{eqnarray}
[r + (\frac{\partial_{r} G}{2 G}) r^2 ](\partial_{r} \Phi ) = r^2 [ \partial_{r}^{2} \Phi -  (\frac{\partial_{r} G}{2 G})(\partial_{r} \Phi )]
\label{3.6}
\end{eqnarray}

Let $\partial_{r} \Phi = P (r, t)$. Then (\ref{3.6}) implies
\begin{eqnarray}
\frac{\partial_r P}{P} =  \frac{\partial_r G}{G} + \frac{1}{r}
\label{3.7}
\end{eqnarray}
We integrate this equation and fix the integration constant by demanding that when $G = \frac{1}{1+r^2}$, $\Phi = - \frac{1}{2} \ln (1 + r^2 )$. This yields
\begin{eqnarray}
P = \partial_{r} \Phi = - r G(r, t).
\label{3.8}
\end{eqnarray}

This gauge choice preserves conformal gauge and when the metric (\ref{3.3}) is that of the Witten black hole, this is the the gauge choice under which the Witten black hole is a fixed point of the flow (\ref{3.4}).
Therefore,
\begin{eqnarray}
\nabla_{r} \partial_r \Phi &=& \partial_r (\partial_r \Phi) - (\frac{\partial_r G}{2 G}) (\partial_r \Phi) \nonumber \\
&&= -G - \frac{1}{2} r (\partial_r G ) \nonumber \\
&& = - \frac{1}{(1+r^2 )^2 } - \epsilon [ F + \frac{1}{2} r (\partial_r F) ]
\label{3.9}
\end{eqnarray}

Note, if we think of $\Phi$ as the dilaton, we are not varying the dilaton independently --- the variation in the dilaton from the background value is dependent on the metric variation, and merely to do gauge fixing.

The linear stability problem of the Witten black hole under the flow (\ref{3.4}) can be set
up by just substituting the ansatz (\ref{3.3}) into the flow (\ref{3.4}) and retaining only terms to $O(\epsilon)$. Rather than do this, for completeness, we will first write the standard linearization of the flow (\ref{3.4}).  as follows: The perturbed metric  $g_{ab}^{p} = g_{ab} + \epsilon h_{ab}$, where henceforth $g_{ab} = g_{ab}^{Witten}$ (the background metric). The linearization of the flow (\ref{3.4}) arises from a standard computation of the linearization of the Ricci tensor (derived in many books; see for instance \cite{chow}). Choosing conformal gauge defined by (\ref{3.9}), the linearized flow of the perturbation $h_{ab}$ is

\begin{eqnarray}
\frac{\partial h_{ab}}{\partial t} = \left [ -
(\Delta_{L}  h)_{ab} + \nabla_{a}\nabla_{b} H -  \nabla_{a}(
\nabla^{c}h_{cb}) - \nabla_{b}( \nabla^{c}h_{ca}) + 2 Q_{ab} \right ] .
\label{3.10}
\end{eqnarray}
Here and in what follows, all covariant derivatives are taken with
respect to the background metric $g$. $H = g^{ab} h_{ab}$ is the
trace of the perturbation. From the ansatz (\ref{3.3}), $H = 2 (1+ r^2 ) F$. $Q_{ab}$ is the appropriate piece proportional to $\epsilon$ in the linearization of the gauge piece  $- 2 \nabla_{a} \partial_b \Phi$. From (\ref{3.9}), we can read off that  $Q_{rr} = 2 [ F + \frac{1}{2} r (\partial_r F) ]$, and consequently $Q_{\theta \theta} = r^2 Q_{rr}$.
\begin{eqnarray}
(\Delta_{L} h)_{ab} = - \Delta h_{ab} + 2 R_{~abc}^{d} h_{~d}^{c} +
 R_{a}^{c} h_{bc} +  R_{b}^{c} h_{ac} \label{3.11}
\end{eqnarray}
is the Lichnerowicz laplacian acting on symmetric 2-tensors (all
curvature tensors being those of the background Witten black hole metric).The
convention we follow for the Lichnerowicz laplacian is that of the
physics literature, and differs from the mathematics one by a
negative sign.

All we need in this linearized approximation is the flow of the function $F(r, t)$. To obtain this, we take the trace of the flow (\ref{3.10}).
This yields
\begin{eqnarray}
\frac{\partial H}{\partial t} = [ \Delta H + 2 g^{ab}Q_{ab}]
\label{3.12}
\end{eqnarray}
We now substitute $H = 2 (1+ r^2 ) F$. Recall that the metric and Laplacian that appear in (\ref{3.12}) are those of the background Witten black hole metric. Then, we obtain the following flow for $F$:
\begin{eqnarray}
\frac{\partial F}{\partial t} = 2 \left [ \frac{(1+ r^2)}{2} \frac{\partial^2 F}{\partial r^2} + \frac{(7 r^2 + 1)}{2 r} \frac{\partial F}{\partial r} + 4 F \right ] .
\label{3.13}
\end{eqnarray}

Now, we first want to investigate if there are unstable perturbations of the sort we have considered (radially symmetric) under the flow. We pick the ansatz
\begin{eqnarray}
F(r, t) = e^{2 q t} \tilde F (r)
\label{3.14}
\end{eqnarray}
Here $q > 0$ is a real number. If we are able to find {\em physically reasonable} perturbations $\tilde F (r)$ solving (\ref{3.13}) for some $q > 0$, then these perturbations would be unstable modes under the flow (they would grow exponentially in $t$). A physically reasonable perturbation must have $\tilde F(r)$ bounded (and it must fall off faster than the background Witten black hole metric) --- otherwise we cannot use linearized perturbation theory. Another (related) requirement that could be imposed is that $\int_{0}^{\infty} \tilde F^2 ~dr$ is finite (square integrable perturbation). Similar stability studies, such as that of Gross,Perry and Yaffe \cite{GPY} define a normalizable perturbation as one for which the `energy' of the perturbation,  $\int \tilde F^2 ~dV$ is finite (the integral being on the Witten black hole manifold and $dV$ is the measure with respect to the Witten black hole metric). We substitute the ansatz (\ref{3.14}) into the equation (\ref{3.13}) and investigate whether there are unstable modes for which the perturbation is reasonable from these various points of view (bounded, and either square integrable or having finite energy). If we do find unstable modes, this would conclusively tell us that the Witten black hole is unstable under first-order RG (Ricci) flow. If we do not find any unstable modes of the type (\ref{3.14}), this would be very interesting. However, a caveat is that this would still not completely guarantee stability of the Witten black hole under the flow --- we cannot rule out unstable modes that are not circularly symmetric, or do not satisfy this ansatz, and we could also have some unstable finite norm superposition of non-normalizable modes of the type (\ref{3.14}).

Substituting ansatz (\ref{3.14}) in (\ref{3.13}), we get
\begin{eqnarray}
\frac{(1+ r^2)}{2} \frac{d^2 \tilde F}{dr^2} + \frac{(7 r^2 + 1)}{2 r} \frac{d \tilde F}{dr} + (4-q) \tilde F = 0.
\label{3.15}
\end{eqnarray}

This equation can be solved exactly, and we are very grateful to the referee for pointing this out in an earlier version where we had discussed the approximate solutions to (\ref{3.15}). To solve the equation, we define a new variable $z = - r^2$. In terms of this, the equation (\ref{3.15}) reduces to a hypergeometric equation

\begin{eqnarray}
z(1 - z) \frac{d^2 \tilde F}{dz^2} + (1 - 4 z) \frac{d \tilde F}{dz} - \frac{(4-q)}{2} \tilde F = 0.
\label{3.16}
\end{eqnarray}
From the standard form of the hypergeometric equation
\begin{eqnarray}
z(1 - z) \frac{d^2 \tilde F}{dz^2} + [c - (a+b+1)z] \frac{d \tilde F}{dz} - ab \tilde F = 0,
\label{3.17}
\end{eqnarray}
we can read off that for (\ref{3.16}), the parameter $c = 1$, $(a+b) = 3$, and $ab = \frac{(4-q)}{2} $. Let $q = 2k(k-1)$. Then, since $q > 0$, $k > 1$. \\$ab = - k^2 + k + 2$. We can pick $a = 2 - k$, $b = k+ 1$.

The two linearly independent solutions around $z=0$, denoted by $\tilde F_{1(0)}$ and $\tilde F_{2(0)}$, are ($|z| < 1$)
\begin{eqnarray}
\tilde F_{1(0)} &=& F(a, b; 1; z) \nonumber ;\\
\tilde F_{2(0)} &=& F(a, b; 1; z) \ln z + \nonumber \\& & \Sigma_{n=1}^{\infty} \frac{(a)_{n} (b)_{n}}{n!^2 } z^{n} [\psi(a+n) - \psi(a) + \psi(b+n) - \psi(b) - \nonumber \\ & & 2\psi(n+1) + 2 \psi(1) ].
\label{3.18}
\end{eqnarray}
Here $F(a, b; 1; z)$ is the Gauss hypergeometric function (sometimes written as $F^{2}_{1} (a,b; 1,;z)$), $\psi$ is the Digamma function and the notation $(a)_{n} = a(a+1)...(a+n-1)$ refers to the Pochhammer symbol. The general solution to (\ref{3.16}) is a linear combination of the two solutions in (\ref{3.18}). For the special case $k=2$ ($q=4$), the solution is much simpler, and we will discuss this later.

By replacing $ z = - r^2$, these are the solutions near $r=0$. Clearly $\tilde F_{2(0)}$ is not finite in the limit $r \to 0$ (and not square integrable in a neighbourhood of $r = 0$ with respect to the measure $dr$). So for physical reasons, we will only consider the solution $\tilde F_{1(0)} = F(a, b; 1; z)$.
By a standard linear transformation valid for $|arg(1-z)|< \pi$, we can analytically continue this solution for $|z|>1$. Rewriting in terms of the $r$ variable, the transformation we will consider is
\begin{eqnarray}
F(a, b; 1;~ -r^2) &=& (1+r^2 )^{-a} \frac{\Gamma(b-a)}{\Gamma(b)\Gamma(1-a)} F(a,1-b, a-b+1; \frac{1}{1+r^2}) + \nonumber \\
& & (1+r^2 )^{-b} \frac{\Gamma(a-b)}{\Gamma(a)\Gamma(1-b)} F(b,1-a, b-a+1; \frac{1}{1+r^2}).~~~~~~~~~~~~~~
\label{3.19}
\end{eqnarray}
This is valid except when the Gamma functions appearing above have poles. Let us first assume that $a,b$ are such that the Gamma functions have no poles. Then we see that as $r \to \infty$, $\frac{1}{1+r^2} \to 0$. Substituting for $a$ and $b$ in terms of $k$, since $k > 1$, the leading behaviour of this solution is of the form $(1+r^2 )^{-2 + k}$. Let $k > 2$. Then, clearly the solution diverges, and cannot be considered as a perturbation of the Witten black hole metric - this solution would change the asymptotics of the geometry and we do not consider it. Therefore there are no finite solutions to (\ref{3.15}) when $k > 2$, or equivalently when $q > 4$. When $1 < k < 2$, or equivalently, $0 < q < 4$, the solution does not diverge, but its leading behaviour is $(1+r^2 )^{-2 + k}$ and this falls off asymptotically slower than the Witten black hole metric (which goes to zero as $(1+r^2 )^{-1})$. Thus we cannot consider this solution as a perturbation.

We now consider the cases when the Gamma functions in (\ref{3.19}) have poles, for $k > 1$.  Recall that if $a$ or $b$ is a negative integer, the hypergeometric series for $F(a, b; 1, -r^2)$ terminates to give a polynomial (which is not finite as $r \to \infty$). As can be checked, this includes the cases when $c-a$ and $c-b$ are integers (we assume $k \neq 2$, and will consider this case separately since the solution then simplifies). The last possibility is if $b-a$ is an integer and $c-a$ and $c-b$ are not integers. In this case, as $ r \to \infty$, $F(a, b; 1, -r^2) \sim (r^2)^{-a}$ (see, for instance, \cite{abramowitz}). Again, as before, either the solution is not finite, or it is finite, but changes asymptotics.

For the one possibility we left out, $k=2$ ( $q=4$) --- the equation (\ref{3.15}) can be easily integrated once to give
\begin{eqnarray}
\left [ r (1+ r^2)^3 \frac{d \tilde F}{dr} \right ]  = C_1 .
\label{3.20}
\end{eqnarray}
 $C_1$ is an integration constant. Integrating once more yields
 \begin{eqnarray}
 \tilde F = C_1 ( \frac{1}{4 (r^2 + 1)^2} + \frac{1}{2 (1 + r^2)} + \frac{1}{2} \ln \frac{r^2 }{1 + r^2 } )+ C_2 .
 \label{3.21}
 \end{eqnarray}
 This is not finite everywhere unless $C_1 = 0$. But the constant mode $C_2 $ is not normalizable, and changes the asymptotics of the geometry.

 We conclude that there is no finite solution to (\ref{3.15}) for any $q > 0$ with the right asymptotic fall-off conditions for a perturbation. Thus the Witten black hole has no unstable modes for circularly symmetric perturbations, and is linearly stable under the class of perturbations of the form (\ref{3.14}).

In the next section, we discuss one possible unstable perturbation of the Witten black hole which would be nonnormalizable in a linearized analysis. Discussing such a perturbation would normally not make sense physically. However, there are many studies of specific nonnormalizable modes under world-sheet RG flow (\cite{barbon}, \cite{hw}) and in quantum gravity (\cite{GPY}) which mediate a transition to another geometry, in some cases, accompanied by a topology change. We suggest that this specific nonnormalizable mode leads to geometry change --- to a solution to world-sheet RG flow, the sausage metric \cite{foz}.

\section{A nonnormalizable mode of the Witten black hole}
Consider the metric which is a solution to first order RG flow of the sigma model, defined on the cylinder (of radius $2$),
\begin{equation}
ds^2 = u(x, \bar t ) (d\tilde \theta^2 + dx^2)
\label{4.1}
\end{equation}

where
\begin{equation}
u (x, \bar t) = \frac{\sinh (- \bar t )}{ \cosh x + \cosh \bar t }
\label{4.2}
\end{equation}
$\bar t$ is the flow parameter of the Ricci flow, $x \in \mathbf{R}$ and $\tilde \theta \in \mathbf{R}/4\pi \mathbf{Z}$.\\
This is called the Rosenau metric (\cite{rosenau}) --- and was incidentally found much earlier in the context of sigma models in \cite{foz}. The curvature obtained from this metric is positive, becoming singular at $\bar t=0$ --- so in the metric (\ref{4.2}), $\bar t<0$. We can think of the cylinder topologically as the sphere with the two poles removed. By compactifying the cylinder to a two-sphere (putting back the poles), we can extend this to a metric defined on the (topological) two-sphere (see \cite{chow} for the details). In fact, the distance from the `equator' $x=0$ to any of the poles at $ x = \pm \infty$ is bounded for $\bar t <0 $. Around $x= 0$, taking a limit as $\bar t \to - \infty$, this metric starts to look like a flat cylinder metric - but around the poles, i.e., $x \to \pm \infty$, this is not true. For finite $\bar t$, the geometry is shaped like a sausage (cylinder-like in the middle and like a cigar tip near each of the poles). The sigma model corresponding to this metric, called the sausage model was studied in detail by Fateev, Onofri and Zamolodchikov \cite{foz}. The curvature corresponding to this metric is
\begin{equation}
R = \frac{\cosh \bar t ~\cosh x + 1}{\sinh(- \bar t) (\cosh x + \cosh \bar t )}
\label{4.3}
\end{equation}
Thus the curvature becomes singular as $\bar t \to 0$. Geometrically, the sausage becomes rounder and also shrinks as $\bar t \to 0$. The nonperturbative IR limit of this quantum field theory is therefore nontrivial - possible candidates for this IR limit are discussed in \cite{foz}.\\

It can be shown that a certain backward limit (scaling of $x$ followed by the limit $\bar t \to - \infty $)of this solution yields the Witten black hole (cigar) metric \cite{chow}. Geometrically, this corresponds to zooming in a neighbourhood of the poles of the sausage solution. We first discuss this backward limit, and then suggest, motivated by this, that a specific perturbation of the Witten black hole should lead to the sausage metric.\\

(\ref{4.1}) can be considered as a metric on the plane by removing one of the poles (or adding a point to the cylinder). We define $\theta = \tilde \theta /2$, so that $\theta \in \mathbf{R}/2 \pi \mathbf{Z}$. Further we let $r = e^{x/2}$. Then the metric in these new coordinates is
\begin{equation}
ds^2 =  \frac{8 \sinh(- \bar t)}{ 1 + r^4 + 2 r^2 \cosh \bar t } (dr^2 + r^2 d\theta^2 ).
\label{4.4}
\end{equation}
\\
For small $r$ (i.e., near the pole) and fixed $\bar t$, this metric reduces to a constant multiple of the flat metric. We now do a scaling $r \to re^{\bar t/2}$. This $\bar t$ dependent diffeomorphism is generated by the vector field $\bar X$ with $\bar X^{r} = r/2$, $\bar X^{\theta} =0$. We now get the metric
\begin{equation}
ds^2 =  \frac{8 \sinh(- \bar t) e^{\bar t}}{ 1 + e^{2 \bar t} r^{4} + 2 e^{\bar t} r^{2} \cosh \bar t } (dr^2 + r^2 d\theta^2 ).
\label{4.5}
\end{equation}
\\
Let us now define $ \bar t = 4 t$ and divide the metric (\ref{4.5}) by a factor of $4$. Then the metric
\begin{equation}
ds^2 =  \frac{2 \sinh(- 4 t) e^{4 t}}{ 1 + e^{8 t} r^{4} + 2 e^{4 t} r^{2} \cosh 4 t } (dr^2 + r^2 d\theta^2 )
\label{4.6}
\end{equation}
solves the RG flow equation, with flow parameter $t$ instead of $\bar t$:
\begin{eqnarray}
\frac{\partial g_{ab}}{\partial  t} = - 2
(R_{ab} + \nabla_{a} X_b + \nabla_{b} X_a )
\label{4.7}
\end{eqnarray}
and $(- 2 X)$ is the vector field that generated the scaling $r \to re^{2 t}$  --- because of taking the factor of $(- 2)$ out in the right hand side of (\ref{4.7}), $X^{r} = - r$ and $X^{\theta} =0$. Note that we parametrize the integral curves of $- 2 X$ by $t$, not $\bar t$. Now we consider the metric (\ref{4.6}) in the limit
$t \to - \infty $. We get
\begin{equation}
ds^2 = \frac{1}{ 1 + r^2} (dr^2 + r^2 d\theta^2 )
\label{4.8}
\end{equation}
which is precisely the Witten black hole metric. $X_{r} = g_{rr}X^{r}$, and when $t \to - \infty $, $X_{r} \to \frac{-r}{1+r^2}$. Setting $X_{i} = \nabla_{i} \Phi$, this implies $\Phi \to - \frac{1}{2} \ln (1+r^2)$, which is precisely the dilaton corresponding to the Witten black hole metric. This explicitly shows that after a scaling, the backward (in $t$) limit of the Rosenau metric is the Witten black hole which is a fixed point of (\ref{4.7}).
\\
Now let us find the $t \to - \infty$ limit of (\ref{4.6}) by expanding \emph{to the next order}: we obtain that as $t \to - \infty $,
\begin{eqnarray}
ds^2 \sim  [ \frac{(1 - e^{8t})}{1 + r^4 e^{8t} + r^2 e^{8t} + r^2 } + ......] (dr^2 + r^2 d\theta^2 )
\label{4.9}
\end{eqnarray}
which, when $r^2 e^{8t} < 1$ becomes
\begin{equation}
ds^2 \sim  [ \frac{1}{ 1 + r^2} - e^{8t} + ........ ] (dr^2 + r^2 d\theta^2 )
\label{4.10}
\end{equation}\\
We can view (\ref{4.10}) as a small perturbation of the Witten black hole metric (where the terms denoted by ellipses are smaller than the leading term displayed, when $r^2 e^{8t} < 1$). This is then some expansion in various modes of the form $e^{2qt}$ with $q \ge 4$ --- note that from the results of the previous section, these are precisely those values of $q$ for which no normalizable mode exists. The leading term in the perturbation would then be a constant mode with $q=4$ (referring to the previous section). A constant mode would be a nonnormalizable mode in the sense we discussed in the previous section, but of course, the form of the metric (\ref{4.10}) is only true when $r^2 e^{8t} < 1$. However, taking the difference between the metric in (\ref{4.9}) and the Witten black hole metric, we get a ``perturbation'' that is not normalizable at any finite $t$ (the reason why we write it within quotes). This result leads us to speculate that there is a superposition of nonnormalizable modes of the Witten black hole metric that leads to the sausage metric. Geometrically, this suggests that the perturbation changes the cigar metric asymptotically - curving the cylindrical end. This geometry change is reminiscent of Gross,Perry and Yaffe's nonnormalizable mode for the Schwarzschild instanton in quantum gravity \cite{GPY} which changes the mass at infinity. We can rewrite the above mode in terms of the RG flow parameter by replacing $t = \frac{\alpha'}{2} \tau $.

\section{Discussion} In this paper, we have set up a linearized stability problem for the Witten black hole under world-sheet RG flow. Our conclusion is that there are no unstable modes in a standard linearized mode analysis for circularly symmetric perturbations of the Witten black hole under RG flow. It would be interesting to cast this target-space result as a statement for the deformations of the world-sheet CFT. This stability result may also imply a stability of the black hole in a path integral formulation with the string low energy effective action. We hope to explicitly show this in future work. We also hope to extend our result to perturbations which are not circularly symmetric (in this case, the gauge fixing is more tedious). We also discuss a result from mathematics, which deals with the sausage solution to RG flow. Upon scaling and taking a backward limit in $t$, the solution tends to the Witten black hole. It can be pictured as zooming in on one of the poles of the sausage. However, this leads to a few speculative ideas. In the usual sigma model approach, the target space metric is expanded about a point. This suggests that expanding the sausage sigma model metric about one of the poles should correspond to the sigma model with the scaling described in the previous section. The natural interpretation for the backward limit is the UV limit of the sigma model, which should then be the CFT corresponding to the Witten black hole. Under a specific ``perturbation'' (which we describe in linearized theory as a superposition of nonnormalizable modes), the sigma model would then flow to the sausage - however, the flow would change the asymptotics. We do not have a proof for this set of speculative ideas. In fact, it would be a very hard problem to address in the target space picture. We think this would be an interesting question to attempt in the world-sheet picture, since the black hole CFT is known, and as well, the sausage is believed to be described by an integrable QFT \cite{foz}.

\section{Acknowledgements} We are indebted to the referee for pointing out while reviewing a previous version that (\ref{3.15}) could be solved exactly --- this enabled us to replace the analysis of this equation using various approximations with the exact solution, and thus obtain a more rigorous result in section III. Part of this work was initiated during a summer project when Carolyn Lambert was an undergraduate student at the University of Alberta, and during this time, she was supported by an undergraduate student research award from NSERC of Canada.


\begin{thebibliography}{100}
\bibitem{mandal} G Mandal, A Sengupta, SR Wadia, Mod Phys Lett A6 (1991) 1685.
\bibitem{witten} E Witten, Phys Rev D44 (1991) 314.
\bibitem{dvv} R Dijkgraaf, H Verlinde, E Verlinde, Nucl Phys B371 (1992) 269.
\bibitem{foz} VA Fateev, E Onofri, Al.B Zamolodchikov, Nucl Phys B406 (1993) 521.
\bibitem{rosenau} P Rosenau, Phys Rev Lett 74 (1995) 1056.
\bibitem{HMT} M Headrick, S Minwalla, T Takayanagi, Class Quant Grav 21 (2004)
S1539.
\bibitem{abramowitz} M Abramowitz and IA Stegun, Handbook of Mathematical functions, National Bureau of Standards Applied mathematics series - 55 (1964).
\bibitem{barbon} JLF Barbon, E Rabinovici, Found Phys 33 (2003) 145.
\bibitem{hw} M Headrick, T Wiseman, Class Quant Grav 23 (2006) 6683; see also M Headrick, S Kitchen, T Wiseman, Class.Quant.Grav.27:035002 (2010).
\bibitem{GPY} DJ Gross, MJ Perry, LG Yaffe, Phys Rev D25 (1982) 330.
\bibitem{tseytlin} AA Tseytlin, Phys Rev D75:064024 (2007).
\bibitem{ds} P Daskalopoulos, N Sesum, 	arXiv:math/0603525v2.
\bibitem{marcus} N Marcus and Y Oz, Nucl Phys B407 (1993) 429.
\bibitem{peet} A Peet, L Susskind, L Thorlacius, Phys Rev D48 (1993) 2415; see also BC Georgalas and E Papantonopoulos, Mod Phys Lett A10 (1995) 1277, JD Hayward, Phys Rev D52 (1995) 1019.
\bibitem{gs} J Gegenberg, V Suneeta, JHEP 0609 (2006) 045.
\bibitem{perelman} G Perelman, arXiv:math.DG/0211159, arXiv:math.DG/0303109, arXiv:math.DG/0307205.
\bibitem{cao} H-D Cao, Recent advances in geometric analysis, Adv Lect Math (ALM), 11 (2009) 1-38.
\bibitem{chow} \emph{The Ricci flow: an Introduction}; B Chow and D Knopf, Mathematical Surveys and Monographs Volume 110, American Mathematical Society.

\end{thebibliography}
\end{document}